# FINE SYNCHRONIZATION THROUGH UWB TH-PPM IMPULSE RADIOS

Moez Hizem[1] and Ridha Bouallegue[2]

[1]6'Tel Research Unit, Higher School of Communications of Tunis, Sup'Com, Tunisia
`moezhizem@yahoo.fr`
[2]Sup'Com, Tunisia
`ridhabouallegue@gmail.com`

## ABSTRACT

*In this paper, a novel fine timing algorithm has been tested and developed to synchronize Ultra-Wideband (UWB) signals with pulse position modulation (PPM). By applying this algorithm, we evaluate timing algorithms in both data-aided (DA) and non-data-aided (NDA) modes. Based on correlation operations, our algorithm remains operational in practical UWB settings. The proposed timing scheme consists of two complementary floors or steps. The first floor consists on a coarse synchronization which is founded on the recently proposed acquisition scheme based on dirty templates (TDT). In the second floor, we investigate a new fine synchronization algorithm which gives an improved estimate of timing offset. Simulations confirm performance improvement of our timing synchronization compared to the original TDT algorithm in terms of mean square error.*

## 1. INTRODUCTION

UWB radios have attracted increasing importance in short-range high-speed wireless communications due to their potential to offer high user capacity with low-complexity and low-power transceivers [1]-[3]. Most of these benefits initiate from the distinctive characteristics inherent to UWB wireless transmissions [3]. However, to harness these benefits, one of the most critical challenges is to obtain an accurate timing synchronization and more specifically timing offset estimation. The complexity of which is accentuated in UWB, compared to narrowband systems, owing to the fact that information bearing waveforms are impulse-like and have low amplitude. In addition, the difficulty of timing UWB signals is furthermore induced by the dense multipath channel that remains unknown at the synchronization step. These reasons give explanation why synchronization has obtained so much importance in UWB research [4]-[7].

Typically, pulse position modulation (PPM), pulse amplitude modulation or on/off keying (OOK) is employed. PPM modulation transmits pulses with constant amplitude and encodes the information according to the position of the pulse, while PAM and OOK use the amplitude for this purpose. Moreover, PPM is regularly implemented to reduce transceiver complexity in UWB systems. But unlike pulse amplitude modulation (PAM) applied in the context of UWB systems, the difficulty of accurate synchronization is accentuated in PPM UWB systems owing to the fact that information is transmitted by the shifts of the pulse positions.

In the last years, numerous timing algorithms have been studied for UWB impulse radios under various operating environments (see e.g., [4], [5], [7]-[16]). In [10], a blind synchronization algorithm that takes advantage of the shift invariance structure in the frequency domain is proposed. An accurate signal processing model for a Transmit-reference UWB (TR-UWB) system is given in [11]. The model considers the channel correlation coefficients that can be estimated blindly. In [12], the authors proposed a code-assisted blind synchronization (CABS) algorithm which relies on the discriminative nature of both the time hopping code and a well-designed polarity code. Timing with dirty templates (TDT), which is the starting point of this paper, was introduced in [15] for rapid synchronization and was developed in [16] for PPM-UWB signals with direct sequence (DS) and/or time hopping (TH) spreading. This technique is based on correlating adjacent symbol-long segments of the received waveform. Except [16], all these timing algorithms are developed for PAM-UWB signals. Since their operations greatly rely on zero-mean property of PAM, these presented timing algorithms are not appropriate to PPM-UWB signals.

In this paper, we develop a fine synchronization algorithm for UWB TH-PPM signals. To improve the synchronization performance of the original TDT developed in [16], our contribution will be to implement a new fine synchronization step and insert it after the coarse one, which is original TDT. The principle of our fine synchronization algorithm is to make a fine search to find the exact moment of pulse beginning (fine estimation of timing offset). This is realized by correlating two consecutive symbol-long segments of the received waveform but this time in an interval that corresponds to the number of frames included in one data symbol. Simulation results show that this new synchronizer using TDT can realize a lower mean square error (MSE) than the original TDT in both non-data-aided (NDA) and data-aided (DA) modes.

The rest of this paper is organized as follows. In Section 2, we outline the PPM-UWB system model. Section 3 describes first the TDT algorithm in a single-user links and upper bounds on the mean square error of TDT estimators in both NDA and DA modes are derived. Then, we defined and developed our fine synchronization algorithm. In Section 4, simulations are yielded to corroborate our analysis. Conclusions are given in Section 5.

## 2. SYSTEM MODEL FOR SINGLE-USER LINKS

In UWB impulse radios, each information symbol is transmitted over a $T_s$ period that consists of $N_f$ frames [1]. During each frame of duration $T_f$, a data-modulated ultra-short pulse p(t) with duration $T_p \ll T_f$ is transmitted from the antenna source. With PPM modulation (see e.g., [17], [18]), the transmitted signal in single-user links is described by the following model:

$$v(t) = \sqrt{\varepsilon} \sum_{k=0}^{+\infty} \sum_{i=0}^{N_f-1} p(t - iT_f - c_{th}(i)T_c - kT_s - d_i\delta) \tag{1}$$

where ε is the energy per pulse, $d_i \in (0,1)$ represents the i-th information bit transmitted, and δ is the time shift associated with binary PPM. User separation is realized with pseudo-random TH-codes $c_{th}(i)$, which time-shift the pulse positions at multiples of the chip duration $T_c$ [1]. In this paper, we focus on a single user link and treat multi-user interference (MUI) as noise.

The transmitted signal propagates through the multipath channel with impulse response:

$$g(t) = \sum_{l=0}^{L-1} \alpha_l \delta(t - \tau_l) \tag{2}$$

where $\{\alpha_l\}_{l=0}^{L-1}$ and $\{\tau_l\}_{l=0}^{L-1}$ are amplitudes and delays of the L multipath elements, respectively. The channel is assumed quasi-static and among $\{\tau_l\}_{l=0}^{L-1}$, $\tau_0$ represents the propagation delay of the channel.

Then, the received waveform is given by:

$$r(t) = \sqrt{\varepsilon} \sum_{l=0}^{L} \alpha_l \sum_{k=0}^{+\infty} p_T(t - kT_s - \tau_{l,0} - \tau_0 - d_i\delta) + \eta(t) \qquad (3)$$

where $\tau_{l,0}$ is arbitrary reference at the receiver representing the delay relative to the arrival moment of the first pulse, $\eta(t)$ is the additive noise and $p_T(t)$ denotes the received symbol waveform as:

$$p_T(t) = \sum_{i=0}^{N_f-1} p(t - iT_f - c_{th}(i)T_c) * g(t + \tau_0) \qquad (4)$$

where * indicates the convolution operation. We define the timing offset as $\Delta\tau := \tau_{l,0} - \tau_0$. Let us suppose that $\Delta\tau$ is in the range of $[0, T_s)$ and we will show in the rest of this paper that this assumption will not affect the timing synchronization. Let $p_R(t)$ the overall received symbol-long waveform defined as follows:

$$p_R(t) = \sum_{l=0}^{L} \alpha_l p_T(t - \tau_{l,0}) \qquad (5)$$

Using (5), the received waveform in (3) becomes:

$$r(t) = \sqrt{\varepsilon} \sum_{k=0}^{+\infty} p_R(t - kT_s - \tau_0 - d_i\delta) + \eta(t) \qquad (6)$$

In the next section, we will develop a low-complexity fine synchronization approach using TDT synchronizer in order to find the desired timing offset. It will be evaluated in both non-data-aided (NDA) and data-aided (DA) modes, without acquaintance of the transmitted sequence and the multipath channel.

## 3. FINE SYNCHRONIZATION APPROACH FOR UWB TH-PPM IMPULSE RADIOS

As mentioned previously, our proposed timing scheme consists of two complementary floors or steps. The first is based on a coarse (or blind) synchronization that is timing with dirty templates (TDT) developed in [16]. The second consists of a new fine synchronization algorithm which further improves the timing offset found in the first floor. We will first give an outline of the TDT approach for UWB TH-PPM impulse radios to better understand the overall timing synchronization suggested in this paper.

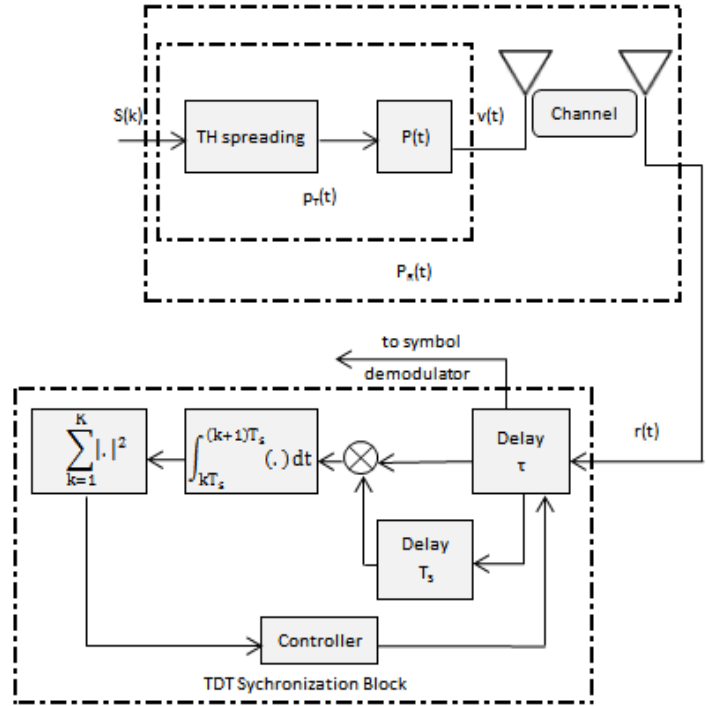

Figure 1. Description of our model with first stage synchronization

### 3.1. TDT for UWB TH-PPM Impulse Radios

The basic idea behind TDT is to find the maximum of square correlation between pairs of successive symbol-long segments. The description of our system model with first stage synchronization (TDT) is illustrated in Fig.1. First, we will start by analyzing $\tilde{\tau}_0$ which represents estimate offset of $\tau_0$ with deriving upper bounds on their mean square error (MSE) in both non-data-aided (NDA) and data-aided (DA) modes.

For UWB TH-PPM systems, a correlation between the two adjacent symbol-long segments $r_k(t) = r(t + kT_s)$ and $r_{k+1}(t) = r(t + (k+1)T_s)$ is achieved [16]. Let $x(k; \tau)$ the value of this correlation $\forall k \in [1, +\infty)$ and $\tau \in [0, T_s)$,

$$x(k; \tau) := \int_0^{T_s} r_{k+1}(t; \tau) \tilde{r}_k(t; \tau) dt$$

$$\tilde{r}_k(t; \tau) := r_k(t + \delta; \tau) - r_k(t - \delta; \tau) \qquad (7)$$

By applying the Cauchy-Schwartz inequality and exploiting the statistical properties of the signal and noise [16], the mean square of the samples in (7) is given by:

$$E_{s,\zeta}\{x^2(k; \tau)\} \approx \frac{1}{2}\left(\varepsilon_R^2 - 3\varepsilon_A(\tilde{\tau}_0)\varepsilon_B(\tilde{\tau}_0) + 2\sigma_\zeta^2\right) \qquad (8)$$

where $\varepsilon_A(\tau) := \varepsilon \int_{T_s-\tau}^{T_s} p_R^2(t)dt$, $\varepsilon_B(\tau) := \varepsilon \int_0^{T_s-\tau} p_R^2(t)dt$, and $\sigma_\zeta$ is the power of $\zeta(k; \tau)$ corresponding to the superposition of three noise terms [16] and can be approximated as an additive white Gaussian noise (AWGN) with zero mean. We notice that $\varepsilon_B(\tilde{\tau}_0) + \varepsilon_A(\tilde{\tau}_0) = \varepsilon \int_0^{T_s} p_R^2(t)dt := \varepsilon_R$ for $\tilde{\tau}_0 \in [0, T_s)$, where $\varepsilon_R$ represents the constant energy corresponding to the unknown aggregate template at the receiver.

Similarly to PAM signals in [15], the term $E_{s,\zeta}\{x^2(k;\tau)\}$ reached its unique maximum at $\tilde{\tau}_0 = 0$. In the practice, the mean square of $x^2(k;\tau)$ is estimated from the average of different values $x^2(k;\tau)$ for k ranging from 0 to M – 1 obtained during an observation interval of duration $MT_s$. In what follows, we summarize the TDT algorithm for UWB TH-PPM systems in its NDA form and then in its DA form.

### 3.1.1. Non-Data-Aided Model for UWB TH-PPM System

For the synchronization mode NDA, the synchronization algorithm is defined as follows:

$$\hat{\tau}_{0,nda} = \arg\max_{\tau \in [0,T_s]} x_{nda}(M;\tau)$$

$$x_{nda}(M;\tau) = \frac{1}{M}\sum_{m=0}^{M-1}\left(\int_0^{T_s} r_k(t;\tau)\tilde{r}_{k-1}(t;\tau)dt\right)^2 \quad (9)$$

The estimator $\hat{\tau}_{0,\,nda}$ in (9) can be verified to be m.s.s. consistent by deriving the mean and variance of the function $x_{nda}(M;\tau)$ [16].

### 3.1.2. Data-Aided Model for UWB TH-PPM System

The blind (NDA) algorithm is principally attractive for its spectral efficiency. Nevertheless, the number of samples M required for reliable estimation can be reduced noticeably if a data-aided (DA) approach is pursued. For this reason, the training sequence $\{s_k\}$ should be considered such that no successive symbols are the same. Therefore, the training sequence for DA TDT is considered to comprise a repeated pattern (for example (1,0, 1,0)); that is:

$$s(k) = \{k+1\}_2 \quad (10)$$

With this pattern, it can be easily verified that the mean square in (8) becomes:

$$E_{s,\zeta}\{x^2(k;\tau)\} = \varepsilon_R^2 - 4\varepsilon_A(\tilde{\tau}_0)\varepsilon_B(\tilde{\tau}_0) + \sigma_\zeta^2 \quad (11)$$

With the NDA approach, it is necessary to take expectation with respect to $s_k$ in order to remove the unknown symbol effects; while the DA mode, this is not needed. Hence, the sample mean $M^{-1}\sum_{m=0}^{M-1} x^2(k;\tau)$ converges faster to its expected value in (11). This pattern is particularly attractive, since it permits a very rapid acquisition which is a major benefit of the DA mode. Data-aided TDT for UWB TH-PPM signals can be accomplished even when TH codes are present and the multipath channel is unknown [16], using:

$$\hat{\tau}_{0,da} = \arg\max_{\tau \in [0,T_s]} x_{da}(M;\tau)$$

$$x_{da}(M;\tau) = \left(\frac{1}{M}\sum_{m=0}^{M-1}\int_0^{T_s} r_k(t;\tau)\tilde{r}_{k-1}(t;\tau)dt\right)^2 \quad (12)$$

The estimator in (12) is essentially the same as (9), except training symbols used in (12). However, theses training symbols are essential in improving the estimation performance. This is approved by the simulation results in Section 4.

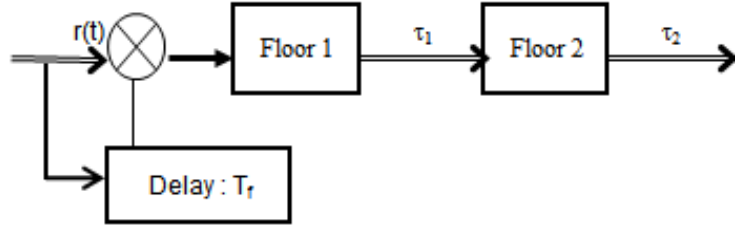

Figure 2. Block diagram of our synchronization scheme

### 3.2. Proposed Fine Synchronization Algorithm

In this part, we propose a low-complexity novel fine synchronization approach using the TDT synchronizer in order to have a more accurate estimate of the exact time synchronization [19]. As mentioned previously, the proposed synchronization algorithm consists of two complementary floors (or steps). The first floor consists of a coarse synchronization, which is none other than the TDT approach presented in the previous part. In the second stage which will be studied, we develop a fine synchronization algorithm that will give a better estimate of the time delay. The block diagram of our synchronization scheme is shown in Fig 2.

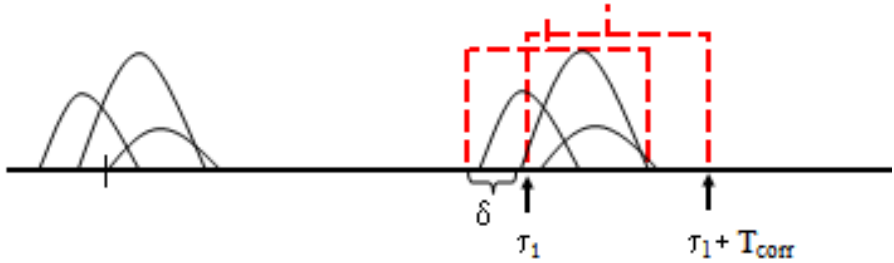

Figure 3. Principle of second synchronization floor

In what follows, we present the second floor or step of our synchronization approach in UWB TH-PPM systems. This second floor realizes a fine estimation of the frame beginning, after a coarse research in the first (TDT approach). The concept which is based this floor is really straightforward. The idea is to scan the interval $[\tau_1 - T_{corr}, \tau_1 + T_{corr}]$ with a step noted $\delta$ by making integration between the received signal and its replica shifted by $T_f$ on a window of width $T_{corr}$. $\tau_1$ being the estimate delay removed after the first synchronization floor and the width integration window value's $T_{corr}$ will be given in Section 5. This principle is illustrated in Fig.3. We can write the integration window output for the $n^{th}$ step $n\delta$ as follows:

$$Z_n = \sum_{k=0}^{K-1} \left| \int_{\tau_1+n\delta}^{\tau_1+n\delta+T_s} r(t-kT_s)r(t-(k+1)T_s)dt \right| \qquad (13)$$

where $n = -N + 1..0..N - 1$, $N = \lfloor T_{corr}/\delta \rfloor$ and K is the number of frames considered for improving the decision taken at the first floor. The value of n which maximizes $Z_n$ provides the exact moment of pulse beginning that we note $\tau_2 = \tau_1 + n_{opt}\delta$. Thus, the fine synchronization is performed. Finally, note that this approach will be applied in both non-data-aided (NDA) and data-aided (DA) modes and will be applicable for any UWB transmission system. We will see later (Section 4) in what mode this approach gives better result compared to those given by the original approach TDT.

## 4. SIMULATION RESULTS

In this section, we will evaluate the performance of our proposed fine synchronization approach in UWB TH-PPM impulse radio systems. The UWB pulse is the second derivative of the Gaussian function with unit energy and duration $T_p \approx 0.8ns$. Simulations are achieved in the IEEE 802.15.3a channel model CM1 [20]. The sampling frequency is $f_c = 50$ GHz. Each symbol contains $N_f = 32$ frames each with duration $T_f = 35$ ns. We used a random TH code uniformly distributed over $[0, N_c - 1]$, with $N_c = 35$ and $T_c = 1.0$ ns. The width integration window value's $T_{corr}$ is 4 ns. The time shift associated with binary PPM is $\delta = 1$ns. Then, the performance of our synchronization approach is tested for various values of M.

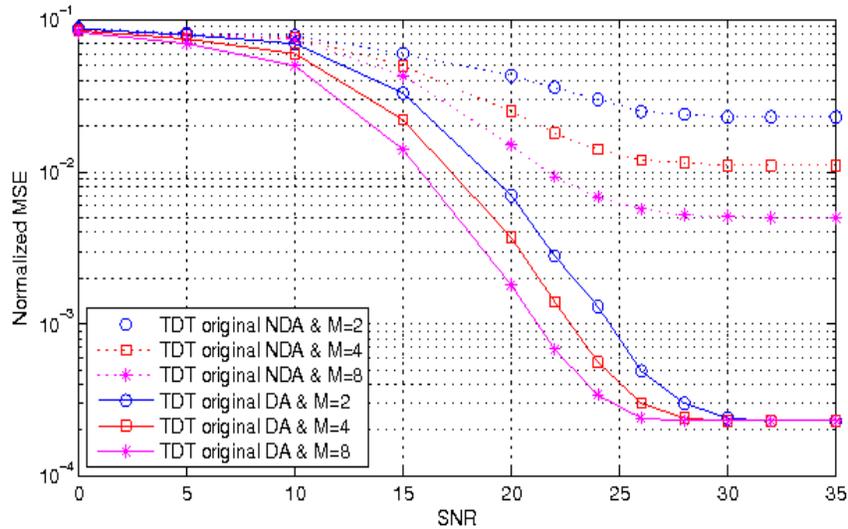

Figure 4. Normalized MSE of original TDT for UWB TH-PPM systems in both NDA and DA modes

In Fig. 4, we first test the mean square error (MSE) of NDA and DA TDT algorithms for UWB TH-PPM systems. The MSE is normalized by the square of the symbol duration $T_s^2$, and plotted versus signal-to-noise rate (SNR). From the simulation results obtained from Fig.4, we note that increasing the duration of the observation interval M leads to improved performance for both NDA and DA modes. We also note that the use of training sequences (DA mode) leads to improved performance compared to the NDA mode.

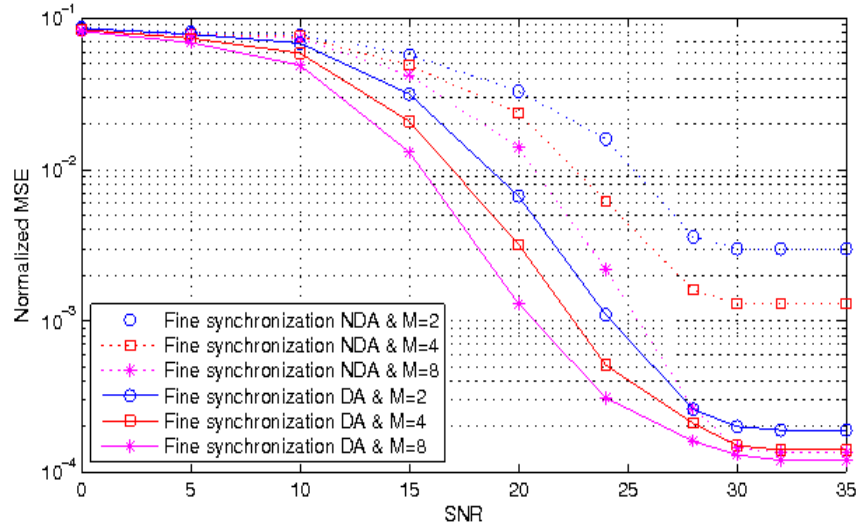

Figure 5. Normalized MSE of our fine synchronization approach for UWB TH-PPM systems in both NDA and DA modes

In Figs. 5-7, we evaluate and compare by simulation the performance of our proposed fine synchronizer with the original NDA and DA TDT algorithms in [16]. For purposes of these simulations, we kept the same channel model and the same TH code parameters as those used previously in Fig. 4. In Fig. 5, we compare the performances of the new fine synchronization approach in both NDA and DA modes.

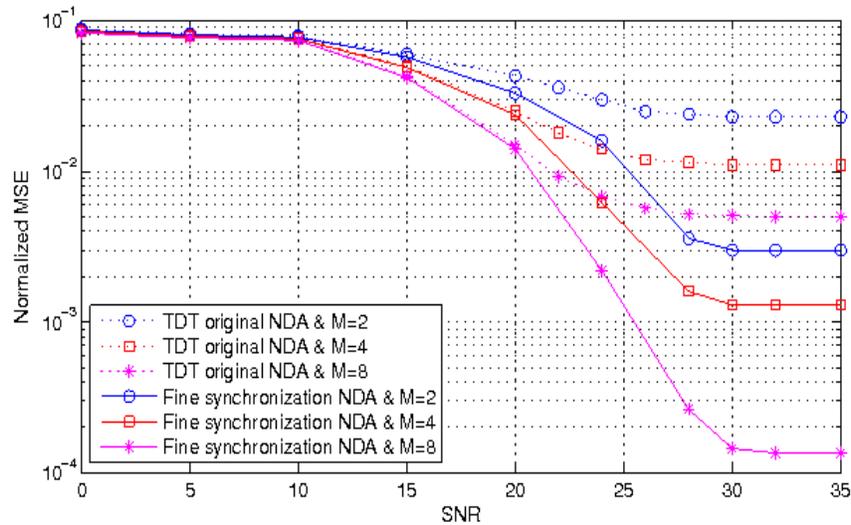

Figure 6. Performances comparison between original TDT and our fine synchronization approaches for UWB TH-PPM systems in NDA mode

In Figs. 6-7, we compare the performances of both original TDT and fine synchronization approach proposed in both NDA and DA modes for different values of M. In comparison with the original TDT approach, we note that the new approach outperforms the NDA mode and offers a slight improvement in DA mode. Even without any training symbol sequence, our synchronizer can greatly outperform the original NDA TDT especially when M is small. This

57

performance improvement is enabled with fine synchronization approach introduced in second floor which can further improve the timing offset found in first floor (coarse synchronization approach : TDT).

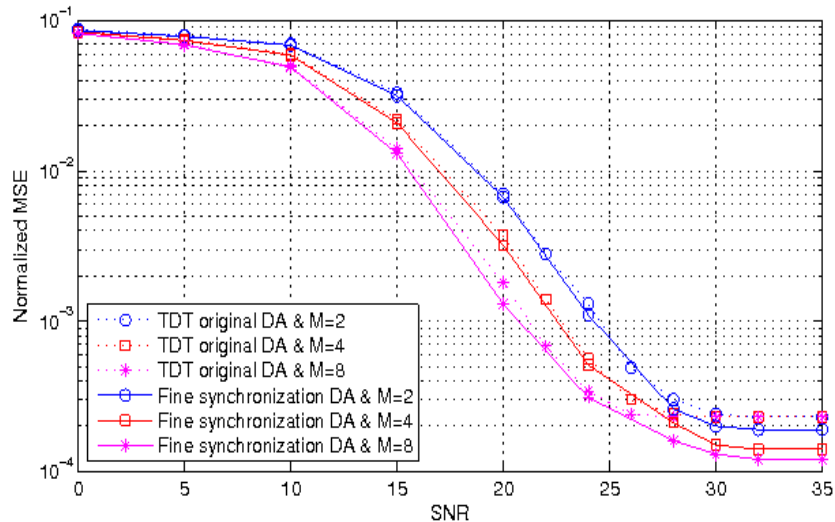

Figure 7. Performances comparison between original TDT and our fine synchronization approaches for UWB TH-PPM systems in DA mode

## 5. CONCLUSIONS

In this paper, we propose a novel fine synchronization scheme using TDT algorithm for UWB TH-PPM impulse radio systems in single-user links. With the fine synchronization algorithm introduced in second floor after TDT (first floor), we can achieve a fine estimation of the frame beginning. The simulation results show that even without training symbols, our new synchronizer can enable a better performance than the original TDT in NDA mode especially when M is small and offers a slight improvement in DA mode.

**Authors**

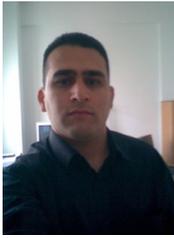

**Moez Hizem** (M'06) received the M.S degree in Electronics in 2004 from the Science Faculty of Tunis (FST), Tunisia and the M.Sc. degree in Telecommunications in 2006 from the National Engineer School of Tunis (ENIT), Tunisia. Since September 2007, he was an university assistant in the High Institute of Technology in Communications at Tunis (ISET'com), he has taught courses in telecommunications and logical systems. He is currently working toward the Ph.D. degree in Telecommunication systems at the High School of Telecommunication of Tunis (SUP'com) in the Laboratory research of System Telecommunication (6'Tel).His current research interests include Wireless systems, Modulation formats and Ultra Wideband Systems.

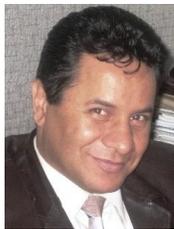

**Ridha Bouallegue** (M'98) received the Ph.D degrees in electronic engineering from the National Engineering School of Tunis. In Mars 2003, he received the Hd.R degrees in multiuser detection in wireless communications. From September 1990 he was a graduate Professor in the higher school of communications of Tunis (SUP'COM), he has taught courses in communications and electronics. From 2005 to 2008, he was the Director of the National engineering school of Sousse. In 2006, he was a member of the national committee of science technology. Since 2005, he was the laboratory research in telecommunication Director's at SUP'COM. From 2005, he served as a member of the scientific committee of validation of thesis and Hd.R in the higher engineering school of Tunis. His current research interests include wireless and mobile communications, OFDM, space-time processing for wireless systems, multiuser detection, wireless multimedia communications, and CDMA systems.